\documentclass[12pt]{spieman}  
\usepackage{amsmath,amsfonts,amssymb}
\usepackage{graphicx}
\usepackage{setspace}
\usepackage{tocloft}
\usepackage[linesnumbered,ruled]{algorithm2e}
\usepackage[noend]{algpseudocode}
\SetKwInOut{Parameter}{Functions}

\makeatletter
\def\BState{\State\hskip-\ALG@thistlm}
\makeatother

\title{DNA Sequence Alignment by Window based Optical Correlator}

\author[a]{Fereshte Mozafari}
\author[b]{Hossein Babashah}
\author[a,*]{Somayyeh Koohi}
\author[b]{Zahra Kavehvash}
\affil[a]{Sharif University of Technology, Department of Computer Engineering, Azadi Ave., Tehran, IRAN, 11155-9517}
\affil[b]{Sharif University of Technology, Department of Electrical Engineering, Azadi Ave., Tehran, IRAN, 1155-4363}

\cftpagenumbersoff{figure}
\cftpagenumbersoff{table} 
\begin{document} 
	\maketitle
	
	\begin{abstract}
		In genomics, pattern matching against a sequence of nucleotides plays a pivotal role  for DNA sequence alignment and comparing genomes. This helps tackling some diseases, such as cancer in humans. The complexity of searching biological sequences in big databases has transformed sequence alignment problem into a challenging field of research in bioinformatics. A large number of research has been carried to solve this problem based on electronic computers.  The required extensive amount of computations for handling this huge database in electronic computers leads to vast amounts of energy consumption for electrical processing and cooling. On the other hand, optical processing due to its parallel nature is much faster than electrical counterpart at a fraction of energy consumption level and cost. In this paper, an algorithm based on optical parallel processing is proposed that not only locate similarity between sequences but also determines the exact location of edits. The proposed algorithm is based on partitioning the read sequence into some parts, namely, windows, then, computing their correlation with reference sequence in parallel. Multiple metamaterial based optical correlators are used in parallel to optically implement the architecture. Design limitations and challenges of the architecture are also discussed in details. The simulation results, comparing with the well-known BLAST algorithm, demonstrate superior speed, accuracy, and much lower power consumption.
	\end{abstract}
	
	
	{\noindent \footnotesize\textbf{*} Somayyeh Koohi,  \linkable{koohi@sharif.edu} }
	
	\begin{spacing}{2}   
		
		\section{Introduction}
		\label{sect:intro}  
		Comparing characters of DNA sequences against a database of reference (i.e., consensus) is defined as DNA sequence alignment. High throughput sequencing (HTS) technologies were introduced in 2006 \cite{1}, and the latest iterations of HTS technologies have an ability to read the genome of a human during three days by the cost of ∼ \$1,000 \cite{2}.

		The Illumina/Solexa sequencing technology \cite{24} typically produces 50-200 million 32-100 base-pairs (bps) reads by only one running \cite{3}. After generating the short reads, they must be mapped (i.e., align) to a known reference genome. The mapping process is computationally very expensive, since the reference genome is very large (e.g., the human genome has 3.2 G bps). So, mapping large volume of short reads to a genome as large as a human genome is a serious challenge for the existing sequence alignment programs.

		A mapper is the name of one software that performs the mapping. The mapper has to search a very large reference genome database to map millions of short reads. On the other hand, each short read may possess edits defined as the difference between bps from read and reference fragment, representing either deletions, insertions, or mismatches, which necessitates expensive approximate searching. To simplify searching a large database, such as a human genome, various studies have developed several algorithms for mappers \cite{4,36}.

		Considering electrical implementation, several mappers have been developed over the past few years, which can be classified based on their mapping algorithms into two categories ; $1)$ hash table based or seed-and-extend mappers similar to the well-known BLAST \cite{5} method, such as mrFAST/mrsFAST \cite{6,7}, MAQ \cite{8}, SHRiMP \cite{9}, Hobbes \cite{10}, drFAST \cite{11} and RazerS \cite{12}; and $2)$ suffix-array or genome compression based mappers that use the Burrows-Wheeler Transform (BWT) such as BWA \cite{3}, Bowtie \cite{14}, and SOAP2 \cite{15}. Each category of mapping algorithms has its own strengths and weaknesses. For performance evaluation of different mappers in finding best alignments, three general metrics are considered: speed of alignment algorithm, accuracy of the algorithm in aligning reads including multiple edits, and finally, comprehensiveness in searching for all aligning locations across the reference genome. The hash table-based mappers compared to the suffix-array based mappers are much slower, more accurate, more comprehensive, and more robust to sequence edits and genomic diversity. For these reasons, hash table based mappers for comparing the different species genomes, are typically preferred, such as mapping reads from a gorilla genome by the human reference genome \cite{24}. On the other hand, suffix-array based mappers, with the BWT optimization, benefits high mapping speed, up to 30-fold faster than hash table based mappers, while by increasing edit distance between the read and the reference fragment, both their mapping accuracy and comprehensiveness reduce. In single nucleotide polymorphism (SNP) discovery studies, where accuracy is less important, suffix-array based mappers are preferred because of their speed \cite{4}. All the aforementioned electronic based mapper algorithms suffer from upper electronic speed limit of transistors. Therefore, it would be suitable to design an architecture that utilizes parallel processing resulting in speed increment.

		Many parallel processing architectures have been proposed including FPGA \cite{16}, GPU \cite{2} and Optics \cite{25,26}. Performance of electronic-based computing, especially in the case of big data processing, is usually limited by high power consumption and inevitably low speed of serial processing \cite{37}. Whereas, optical computing benefits from parallel processing inherited in optics and low power loss such as methods that presented in \cite{27}. Accordingly, among all of the aforementioned approaches, optics is the only solution that can utilize parallelism with no limit and also benefits low power consumption.

		Although optics offers high speed processing as a result of parallel processing, it suffers from some challenges due to its implementation considerations. Specifically, since any arbitrary algorithm cannot be efficiently implemented in optics, any algorithm introduced for sequence alignment should meet its limitations and benefit from the nature of optical processing. Therefore, electronic algorithms and architectures cannot be roughly implemented in optics, making it necessary to present an appropriate algorithm for optical implementation. However, traditional optical components are usually bulky and have slow response \cite{25}. Therefore, new optical components that benefit from small size and fast response, namely metamaterial based components, are of the great interest \cite{35}. Metamaterial based components are artificially engineered structures that can manipulate the impinging light in a sub-wavelength regime and perform computation. In \cite{25}, the concept of computational metamaterial is introduced to perform optical mathematical operations, including spatial differentiation, integration, and convolution. In this approach, manipulating the impinging wave to the desired output is obtained as the wave propagates through the metamaterial structure in order to carry out optical computing process . Finally, this paper proposes a method which not only supports high accuracy and comprehensiveness, but also overcomes electronics speed limitation by using optical parallel processing in less power consumption as well.

		So far, various approaches such as optical correlation \cite{22} and Moir{\'e} matching techniques \cite{34,29} have been proposed to perform DNA sequence alignment by the means of optical computing process. In correlation based methods \cite{22,21,31,30}, DNA sequence alignment is performed through an optical correlator to detect regions with high score of similarity (referred to as global alignment). However, detecting the exact location of edits (referred to as local alignment) which is very critical in the field of bioinformatics \cite{5}, is not addressed by optical correlators. On the other hand, through the Moir{\'e} based matching techniques \cite{34}, the authors have attempted to exploit edit detection based sequence alignment algorithms. Although the proposed algorithm offers fairly acceptable performance for detecting edit locations, its applicability to big data analysis is not addressed. To address above limitations, in this paper, we propose an optical DNA alignment algorithm capable of both global and local alignments, as well as big data analysis.

		In this paper, computational metamaterials are used to perform optical DNA sequence alignment based on optical correlator. For increasing the accuracy of global alignment and detecting exact location of edits, the read sequence is divided into several overlapped parts which are called windows. Implementing correlator with metasurfaces (i.e. ultrathin metamaterial) is performed in the Fourier domain by applying a transfer function proportional to Fourier transform of the reference genome sequence to the impinging optical signal of query genome sequence Fourier transform \cite{25}. The Fourier transformation is also carried out with an input GRIN lens while a metasurface is placed in its focal plane. The analysis for various edit location detection is also considered and the performance of our method is compared against the state of the art electronic based architectures \cite{3,5}. The simulation results demonstrate that the proposed method is promising for DNA sequence alignment operations. In summary, this paper makes the following contributions:
		\begin{itemize}
			\item Proposes a high speed optical correlation algorithm capable of accurate global alignment and locating edits by considering different window sizes.
			\item Proposes a correlation-based algorithm that its implementation is applicable in optics.
			\item Provides an optical structure for the new algorithm.
			\item Gives a comparison of the proposed optical architecture with the well-known BLAST sequence alignment algorithm. The comparison addresses the run-time, hardware complexity, and accuracy of proposed method against those of BLAST mechanism.
			
		\end{itemize}
		The rest of this paper is structured as follows: Section 2 presents the proposed algorithm including details of alignment and detecting the exact locations of edits, as well as the proposed optical structure. Section 3 discusses the simulation results. Finally, the paper is concluded in Section 4.
		\section{Proposed Parallel Optical DNA Sequence Alignment Method}
		For decades, the alignment methods of Dayhoff \cite{17}, Smith-Waterman \cite{18}, and Needleman-Wunsch \cite{19} have been enhanced and refined. Although faster methods, such as FASTA \cite{20} and BLAST \cite{5} algorithms, exist, they generally impose tradeoffs between high speed on one hand, and decreased accuracy or requirement for preprocessing of sequence dataset on the other hand. To overcome the aforementioned limitations, this paper proposes a novel sequence alignment approach based on optical cross-correlation, as a fast process. It is worth noted that the proposed method is robust in dealing with character insertions, deletions, and substitutions. Furthermore, in this paper, the optical cross correlation operation is implemented using multiplication of two sequences in Fourier domain. Hence, this operation, to be performed in the optical domain, benefits from high-speed optical processing. Moreover, the speed up occurs as the throughput increases taking advantage of parallelism inherent in optical processing \cite{28}. The details of the algorithm and sample applications are given in the following subsections.
		\subsection{The Key Algorithm}
				\begin{figure}[!t]
					\begin{center}
						\begin{tabular}{c}
							\includegraphics[height=5.5cm]{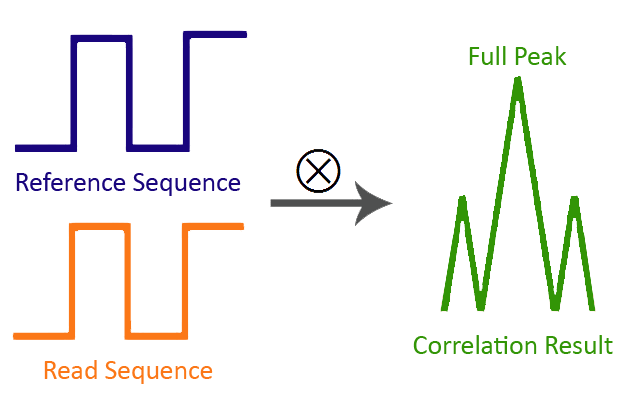}  
						\end{tabular}
					\end{center}
					\caption 
					{ \label{fig1}
						Correlation operation for two signals and correlation result with their corresponding full peak.} 
				\end{figure} 
		A DNA sequence, namely strand, consists of an ordered listing of four nucleic acid bases; represented in the database by the letters G (guanine), C (cytosine), A (adenine), and T (thymine). Thus, a typical sequence might appear as 'AACGTGCCTGG'. In this paper, a DNA sequence is considered as a one-dimensional signal. The alignment problem can then be solved by cross-correlating two signals as follows:
		\begin{equation}
		\label{eq:1}
		c(m) = \sum\limits_n {x(n)y(n - m)} 
		\end{equation}
		where, $x$ and $y$ represent the sequences to be compared, and c defines the cross-correlation function computed by shifting $y$ sequence $m$ units with regard to the $x$ sequence. As shown in Fig. \ref{fig1}, correlation yields a maximum value (referred to as full peak) when two signals are identical with respect to a shift. 
		
\begin{figure}[!t]
			\begin{center}
				\begin{tabular}{c}
					\includegraphics[height=4.5cm]{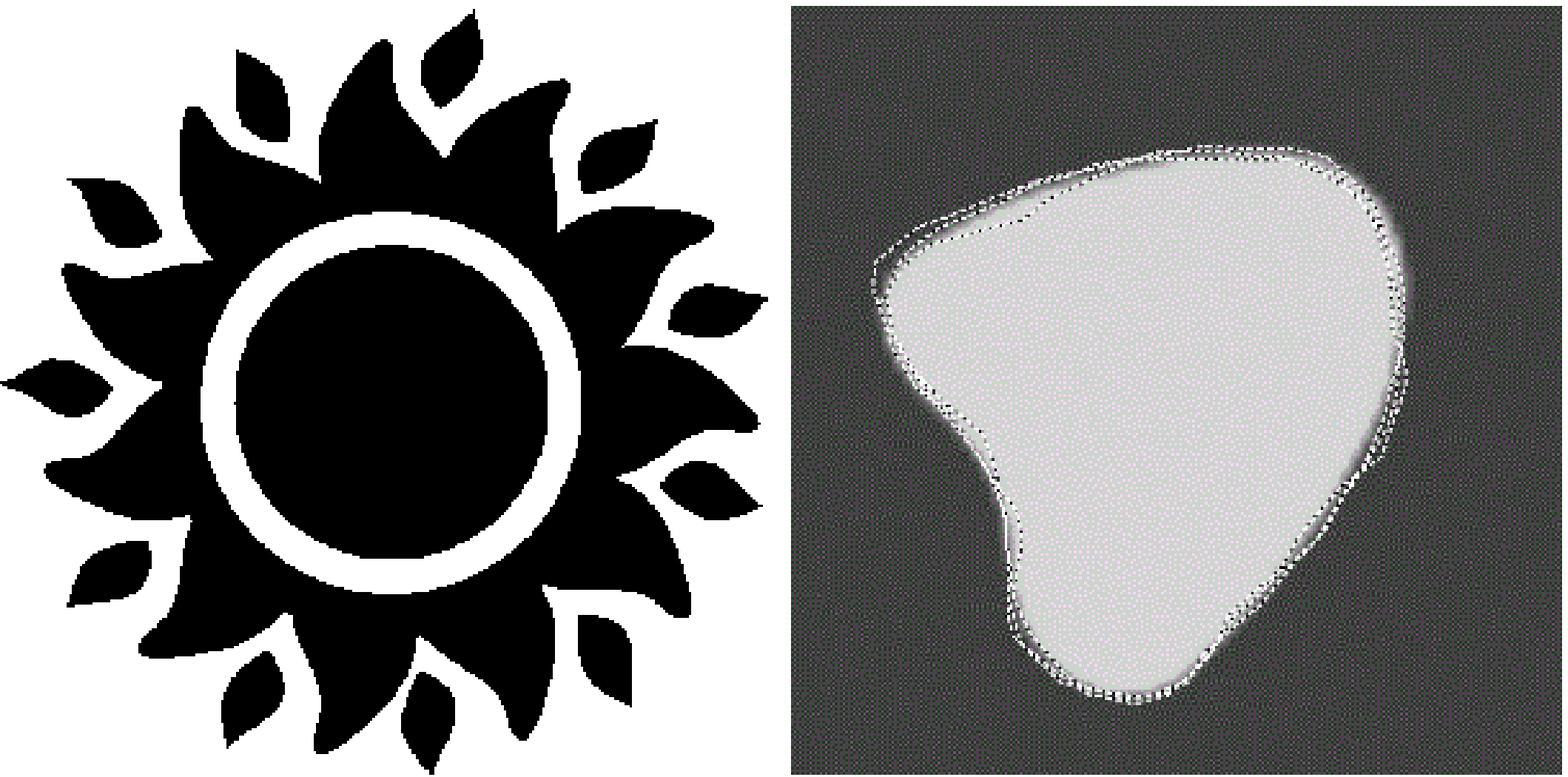}  
				\end{tabular}
			\end{center}
			\caption 
			{ \label{fig2}
				Signal representation of DNA sequence ‘ACCGT’ according to the proposed coding scheme in Table 1.} 
		\end{figure}

		To efficiently perform cross-correlation based sequence alignment algorithm, two properties need to be considered for DNA coding. First, ambiguous matches must be avoided. Second, the number of bits required per base pair should be minimized. The latter property decreases the required computation time and system complexity. For this purpose, as shown in Table \ref{table1}, a suitable code is proposed to represent each DNA base pair (A, C, G, T):\\
		\begin{table}[!h]
			\caption{DNA base pairs coding scheme.} 
			\label{table1}
			\begin{center}       
				\begin{tabular}{|c|c|} 
					\hline
					\rule[-1ex]{0pt}{3.5ex}  Base pair & Code  \\
					\hline\hline
					\rule[-1ex]{0pt}{3.5ex}  A & 1 0 0 0  \\
					\hline
					\rule[-1ex]{0pt}{3.5ex}  C & 0 1 0 0   \\
					\hline
					\rule[-1ex]{0pt}{3.5ex}  G & 0 0 1 0   \\
					\hline
					\rule[-1ex]{0pt}{3.5ex}  T & 0 1 0 0  \\
					\hline
				\end{tabular}
			\end{center}
		\end{table} 
		\\
		As an example, according to the proposed coding scheme, the DNA sequence ‘ACCGT’ is represented by the binary signal shown in Fig. \ref{fig2}.

		\subsection{Problem Definition}
		\begin{figure}[t!]
			\begin{center}
				\begin{tabular}{c}
					\includegraphics[height=1.7cm]{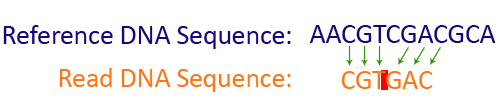}  
					
				\end{tabular}
			\end{center}
			\caption
			{ \label{fig2new}
				Reference and read DNA sequence for comparison.} 
		\end{figure}		
		So far, various optical DNA alignment approaches \cite{29,30,31,newspie1} have been proposed to perform correlation on two signals or images for detecting exact match among DNA strings. Although there might be some edits in one sequence that mess up exact matching, there still exist a large similarity among DNA strings that must be founded. Unfortunately, existing correlation methods \cite{21,22} are mostly unable to find this similarity among sequences at the presence of edits, which is highly desirable. To overcome this limitation, this paper proposes a novel correlation-based matching algorithm to detect DNA sequences similarity and to exactly locate the edits.
		For more clarity, as follows, we explore the correlation-based alignment method through an example. Assume that reference and read sequences to be compared are considered as shown in Fig. \ref{fig2new}.		
		\begin{figure}[t!]
			\begin{center}
				\begin{tabular}{c}
					\includegraphics[height=5.3cm,width=11cm]{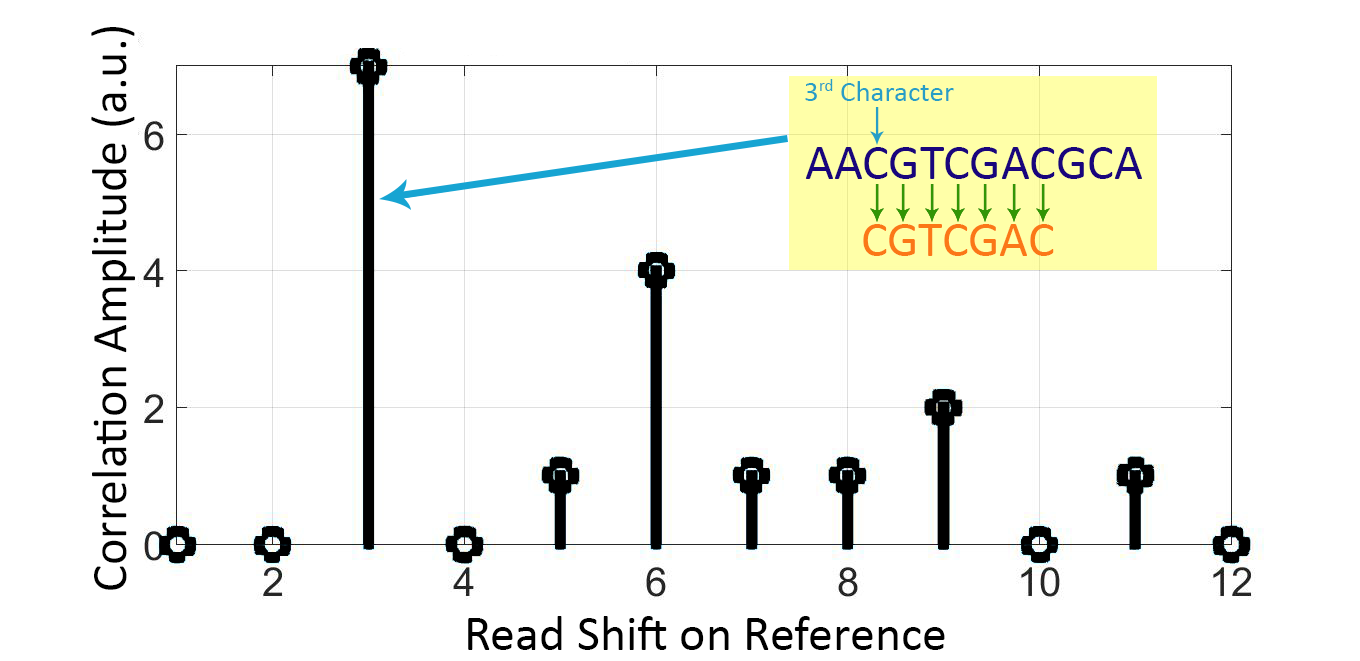} \\
					(a)
					\\
					\includegraphics[height=5.3cm,width=11cm]{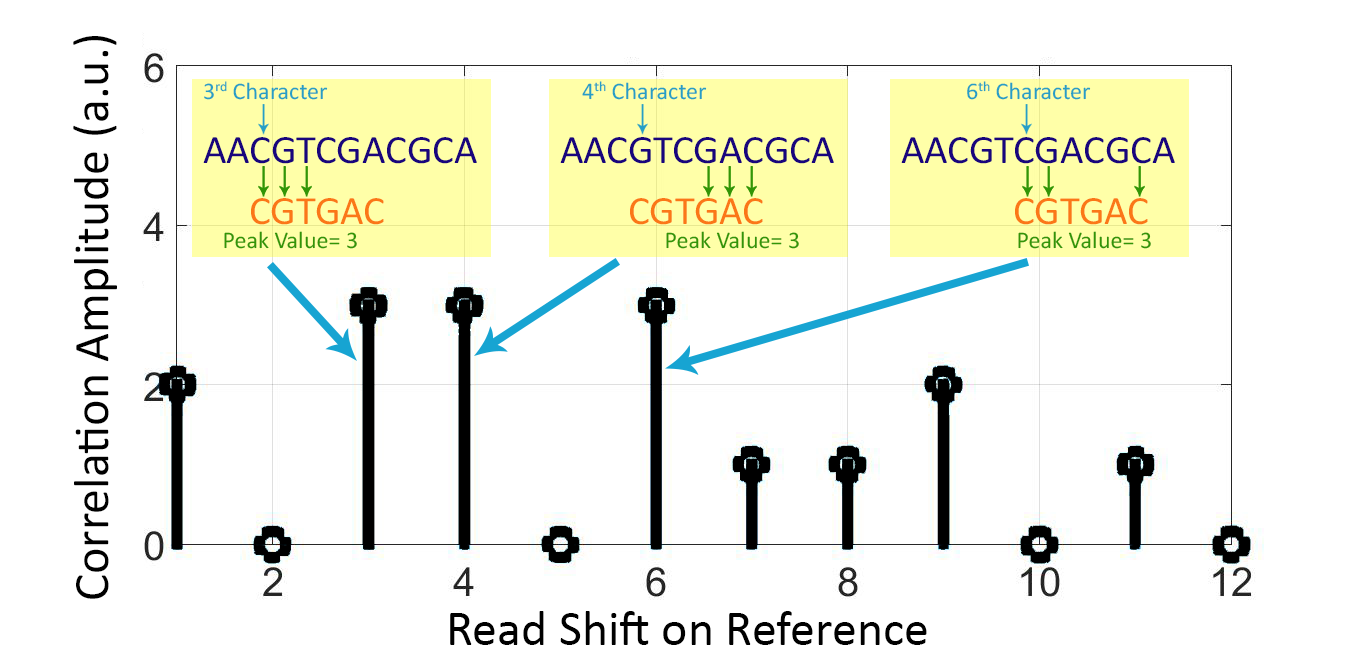} 
					\\
					(b)
				\end{tabular}
			\end{center}
			\caption 
			{ \label{fig3}
				Correlating two sequences. (a) Correlation output prior to character deletion in the read sequence. (b) Correlation output once character deletion is employed in the read sequence.} 
		\end{figure} 
		In fact, the $4^\text{th}$ character of read sequence (‘C’) has been deleted. To illustrate the limited capability of correlation approach for finding sequence mismatches in the case of high similarity among input DNA sequences, first, sequences are coded using the proposed binary digits for each character, then, the reference sequence is correlated by the read sequence before and after ‘C’ character deletion, finally, down sampling is applied to prevent ambiguous matches. The corresponding correlation outputs are depicted in Fig. \ref{fig3}.  As shown in Fig. \ref{fig3}(a), prior to character deletion, there exists high similarity between read and reference sequences from third character of the reference. In this regard, peak value of 7 is obtained, since whole read sequence by the length of 7 characters is exactly matched through its corresponding location in the reference sequence. The full peak, resulted from the exact match, is  determined in Fig. \ref{fig3}(a). On the other hand, Fig. \ref{fig3}(b) presents result of the correlation between reference and read sequences once character deletion is employed. As shown in this figure, three different peaks with the same value of 3 are placed in the third, fourth and sixth location of the reference genome. Therefore, correlating reference and read sequences cannot effectively locate probable character deletion, insertion, and substitution. It is worth noting that some parts of the read sequence might completely match with the reference sequence both prior and after edits, as it can be seen in Fig. \ref{fig3new}.
		\begin{figure}[t!]
			\begin{center}
				\begin{tabular}{c}
					\includegraphics[height=1.9cm]{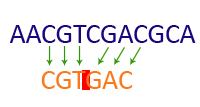}  
				\end{tabular}
			\end{center}
			\caption
			{ \label{fig3new}
				DNA Sequence Alignment for two strings.} 
		\end{figure}
		
		The most remarkable achievement to emerge from the above discussion is that partitioning the read sequence provides advantages resulting in better comparison. With this in mind, a correlation-based sequence alignment algorithm based on partitioning the read sequence is discussed in the next subsection.

		\subsection{Proposed Correlation-Based Alignment Algorithm}
\begin{figure}[!t]
			\begin{center}
				\begin{tabular}{c}
					\includegraphics[height=4.2cm]{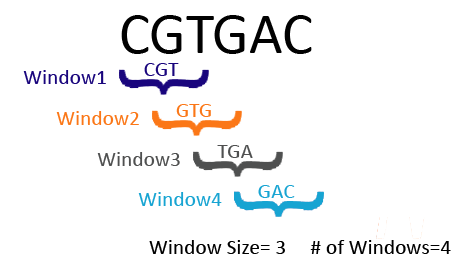}  
				\end{tabular}
			\end{center}
			\caption 
			{ \label{fig4}
				Partitioning read sequence into multiple windows.} 
		\end{figure}

\begin{figure}[!b]
			\begin{center}
				\begin{tabular}{c}
					\includegraphics[height=7.5cm]{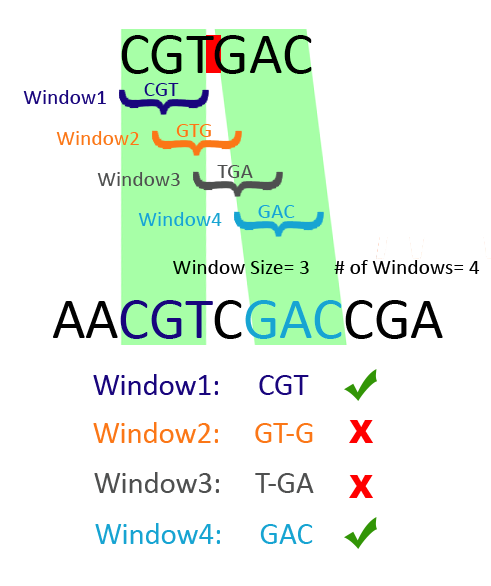}  
				\end{tabular}
			\end{center}
			\caption 
			{ \label{fig5}
				Edit-free windows of read sequence that can be aligned with reference sequence.} 
		\end{figure}

\begin{figure}[!t]
			\begin{center}
				\begin{tabular}{c}
					\includegraphics[height=7.5cm]{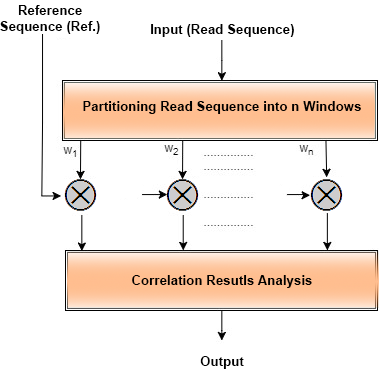}  
				\end{tabular}
			\end{center}
			\caption 
			{ \label{fig6}
				Block-diagram of the proposed correlation based DNA alignment. $ \otimes $ indicates correlation operation.} 
		\end{figure}

		To overcome the aforementioned limitation meet by correlating read and reference DNA sequences, a novel matching algorithm based on optical correlation is proposed which takes advantages of read sequence partitioning. These partitions are introduced with a new concept, named as ‘window’ ($w$), in this paper, which motivate us to name the proposed algorithm as Window-based Optical Correlation (WOC). Number of characters in a window, denotes ‘window size’ ($w_s$), is related to the read length ($L_R$) and number of edits ($e$) in the sequence  through the following equation:
		\begin{equation}
		\label{eq:2}
		{w_s} = \left\lfloor {\frac{{{L_R}}}{{e + 1}}} \right\rfloor 
		\end{equation}
		As the first step of the proposed algorithm, each read sequence, with length ($L_R$), is partitioned into $n$ overlapping windows, as shown in Fig. \ref{fig4}.
		The number of partitions, $n$, in Fig. \ref{fig4} is computed as follows:
		\begin{equation}
		\label{eq:3}
		n = {L_R} - {w_s} + 1
		\end{equation}
		Moreover, window size ($w_s$) is computed by dividing read length ($L_R$) by the number of edits ($e$) plus one. When one or more edits exist in the read sequence, it can still be correctly aligned with the reference sequence as long as there exists one edit-free window of the read sequence. In fact, since edits do not exist in all windows, there are some edit-free windows by breaking the read into $n$ windows. For example, as shown in Fig. \ref{fig5}, the sample sequence is partitioned to 4 windows, out of them two are edit-free and two include edit.
		In general, to detect $e$ (number of edits) edits, each read should be partitioned into windows of size ${\frac{{{L_R}}}{{e + 1}}}$. In this manner, it is easy to show that at least one window will be edit-free. Calculating window size by Equation (\ref{eq:2}) results in the maximum value of window size, and therefore, shorter windows are also possible. However, a trade-off exists between the number of required correlations ($n$) and window size ($w_s$), as follows. As represented by Equation (\ref{eq:3}), subtracting window size from the length of read plus one yields the number of windows. As a result, for every read sequence correlation must be done $n$ times. Hence, as the number of windows increases, the number of required correlation operations increases as well, i.e. decreasing $w_s$  leads to an increment in the value of $n$. Summary of the proposed correlation-based algorithm for DNA sequence alignment using sequence partitioning method is illustrated in Fig. \ref{fig6}.
		\subsection{Correlation-Based Matching Algorithm Analysis}
		\label{sec24}
\begin{figure}[!t]
			\begin{center}
				\begin{tabular}{c}
					\includegraphics[height=13cm]{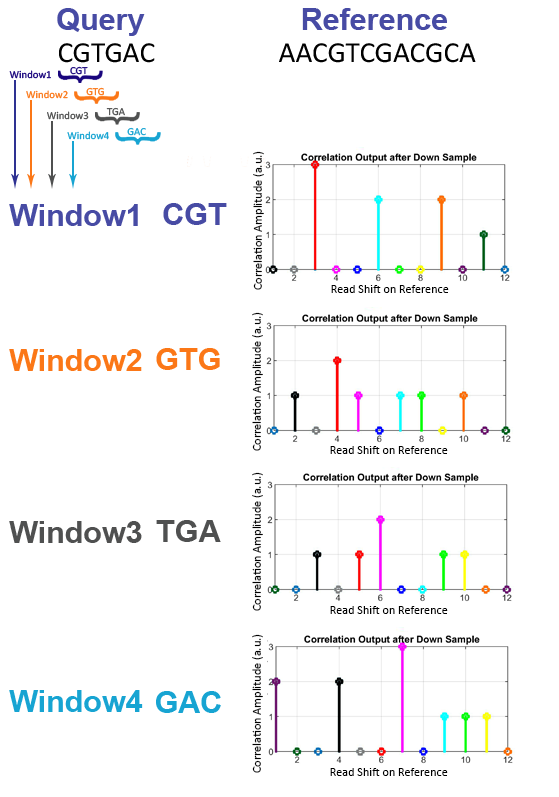}  
				\end{tabular}
			\end{center}
			\caption 
			{ \label{fig7}
				Correlation results for various windows needed to perform global and local alignment.} 
		\end{figure}

		Once the read sequence is partitioned into $n$ windows, each window is correlated with the reference sequence. Fig. \ref{fig7} illustrates the results obtained from correlating all windows of a sample sequence with a reference sequence. 
		However, these plots must be analyzed for detecting edit location to achieve sequence alignment. Sequence alignment is performed into two following steps:
		\begin{enumerate}
			\item Global Alignment:
			
			To perform global alignment, as the first step, each read sequence is partitioned into $n$ windows, then, these $n$ windows are correlated with the reference sequence. The outputs of the corresponding correlations are depicted in Fig. \ref{fig7}. Correlating read sequence with the reference sequence is analyzed through a novel algorithm, explored as follows, considering the value of window size. In this algorithm, read sequence is considered to be located from the first character of the reference sequence and is analyzed by the use of $1^\text{st}$ peak value of $w_1$, $2^\text{nd}$ peak value of $w_2$, $3^\text{rd}$ peak value of $w_3$, and finally $4^\text{th}$ peak value of $w_4$ result. In the same way, for $2^\text{nd}$ character of the reference sequence, $2^\text{nd}$ peak value of $w_1$, $3^\text{rd}$ peak value of $w_2$, $4^\text{th}$ peak value of $w_3$, and $5^\text{th}$ peak value of $w_4$ result are considered, respectively. The latter method of computation is repeated to the end of reference sequence. Aforementioned peak values, that are supposed to be analyzed together, are plotted with the same color in Fig. \ref{fig7}.	
			
			As the second step, total number of full peaks of the same color, shown in Fig. \ref{fig8}, are extracted, which is a number in the range of $0$ to $n$. Actually, this value represents the number of edit-free windows that are exactly matched. Hence, high values of this number represent high similarity. Moreover, analyzing full peaks, we can determine boundary of similar regions.
			
			As shown in Fig. \ref{fig8}, there exist two full peaks; the first one by the beginning of read sequence from third character, and the other one by the beginning of read sequence from forth character of the reference sequence. As a result, there is a global alignment in the range of [$3$, $4+6$]. The later interval is computed by considering locations of full peak plus read length.
			\begin{figure}[!t]
				\begin{center}
					\begin{tabular}{c}
						\includegraphics[height=5.3cm,width=11cm]{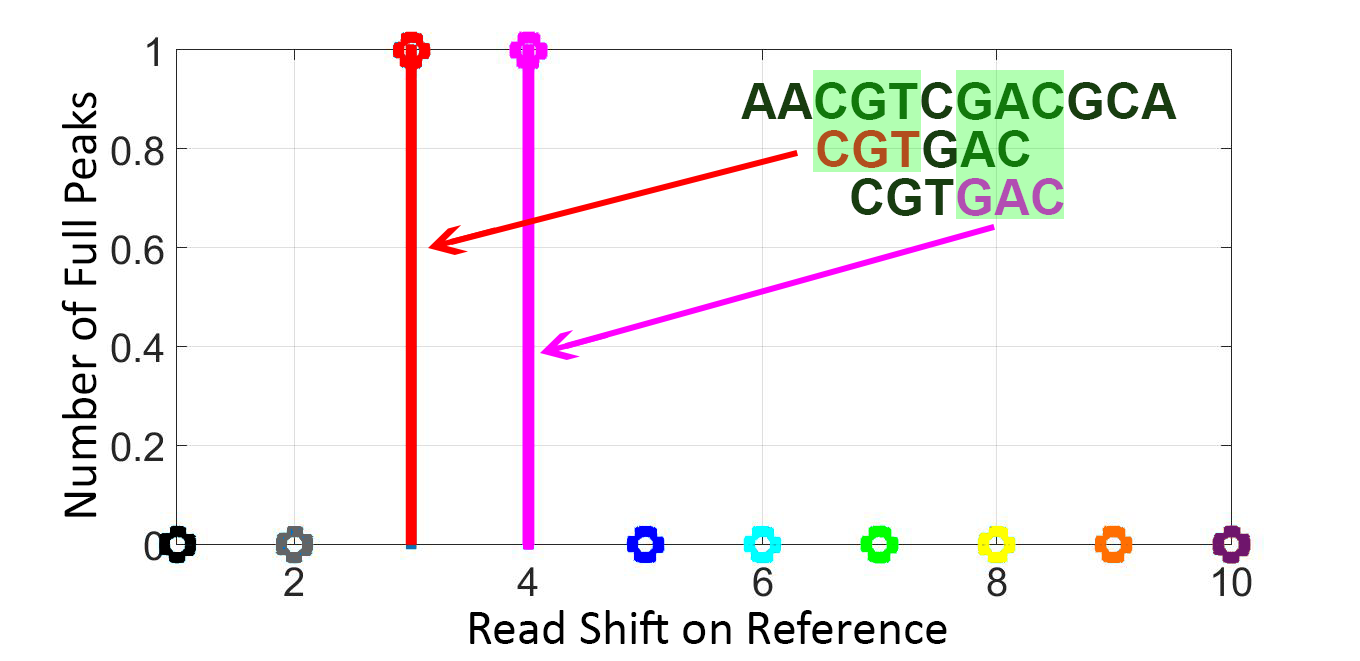}  
					\end{tabular}
				\end{center}
				\caption 
				{ \label{fig8}
					Number of full peaks which is achieved through all windows representing global alignment.} 
			\end{figure} 
			\item Local Alignment:
			
			Once interval of global match has been specified in the reference sequence (e.g. [$3$, $10$] in the aforementioned example), for detecting edit locations, all peaks in this interval are extracted from the correlation results, shown in Fig. \ref{fig7}. Result of thresholding is depicted in Fig. \ref{fig9}, where non-full peaks are shown by zero values.  
			\begin{figure}[!t]
				\begin{center}
					\begin{tabular}{c}
						\includegraphics[height=5.3cm,width=11cm]{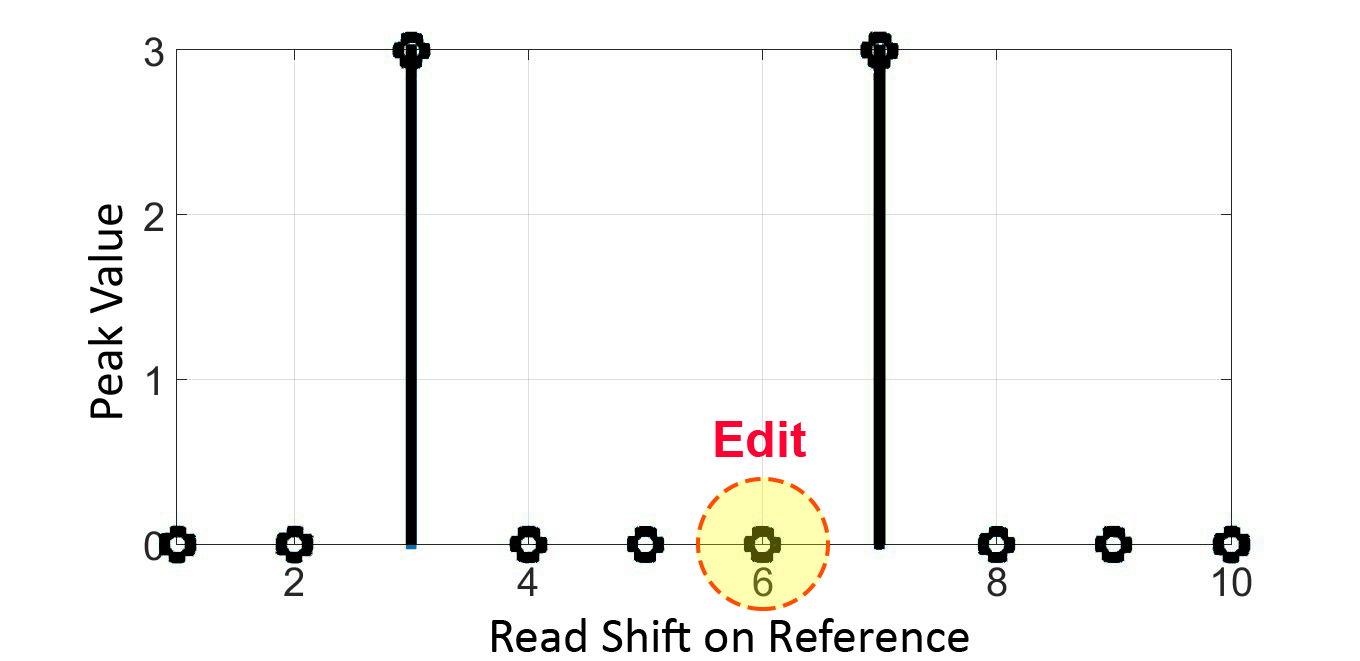}  
					\end{tabular}
				\end{center}
				\caption 
				{ \label{fig9}
					Full peaks in the read sequence for performing DNA sequence local alignment.} 
			\end{figure} 
			Based on Fig. \ref{fig9}, it can be found that one edit (indel) has occurred in the read sequence corresponding to the $6^\text{th}$ character of reference sequence for two reasons; first, there is a full peak at the $3^\text{rd}$ character, which results in no edit at the locations of $3^\text{rd}$, $4^\text{th}$, and $5^\text{th}$ characters (because of $w_s= 3$). On the other hand, since no full peak appears at the forth character so there exists an edit in either $4^\text{th}$, $5^\text{th}$, or $6^\text{th}$ character. Summarizing all above observations, we can conclude that there is an edit in the read sequence corresponding to the $6^\text{th}$ character of the reference sequence. Second, there is full peak and non-full peak at $7^\text{th}$ and $6^\text{th}$ characters, respectively, so there is an indel before $7^\text{th}$ character. All in all, it can be concluded that with respect to the full and non-full peak locations, edit locations will be detected.
		\end{enumerate}
		\subsubsection{Case Study}
		In this section, edit locations are analyzed in a case study including two edits. 
		
\begin{figure}[t!]
			\begin{center}
				\begin{tabular}{c}
					\includegraphics[height=1.9cm]{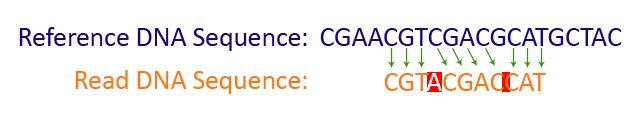}  
				\end{tabular}
			\end{center}
			\caption 
			{ \label{fig10}
				An example considering one insertion and one deletion.} 
		\end{figure}

		Consider the reference and read sequences to be considered as shown in Fig. \ref{fig10}.
					\begin{figure}[!b]
						\begin{center}
							\begin{tabular}{c}
								\includegraphics[height=5.3cm,width=11cm]{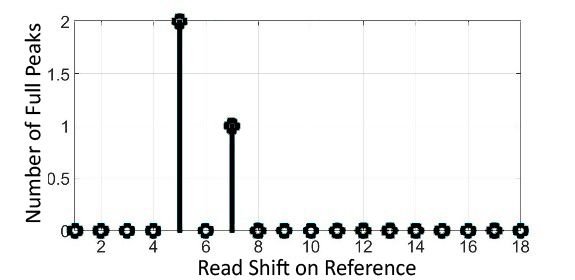}  
							\end{tabular}
						\end{center}
						\caption 
						{ \label{fig11}
							Number of full peaks to perform global alignment obtained from all windows.} 
					\end{figure}  
		In this case study, there is one character insertion and one character deletion, so the viable $e$ equals two. Moreover, according to Equations (\ref{eq:2}) and (\ref{eq:3}), $w_s$ and $n$ equal to $3$ and $10$, respectively. As follows, we investigate how the proposed global and local alignment algorithms exactly locate the indels.
		\begin{enumerate}
			\item Global Alignment:
			
			With respect to the aforementioned global alignment algorithm, discussed in Section \ref{sec24}, the reference sequence is correlated with $10$ windows. In the next step, full peaks are extracted over $10$ correlation window plots, as discussed in Section \ref{sec24}, and the result is shown in Fig. \ref{fig11}. As this figure illustrates, there is a global matching in the range of [$5$,$15$]. 
			\item Local Alignment:
			
			Once interval of global match has been specified to be [$5$,$15$] in this case study, for exactly locating edits in the read sequence, peaks within this interval are extracted from $10$ correlation window plots . As the result is shown in Fig. \ref{fig12}, non-full peaks are shown by zero values.
			\begin{figure}[!t]
				\begin{center}
					\begin{tabular}{c}
						\includegraphics[height=5.3cm,width=11cm]{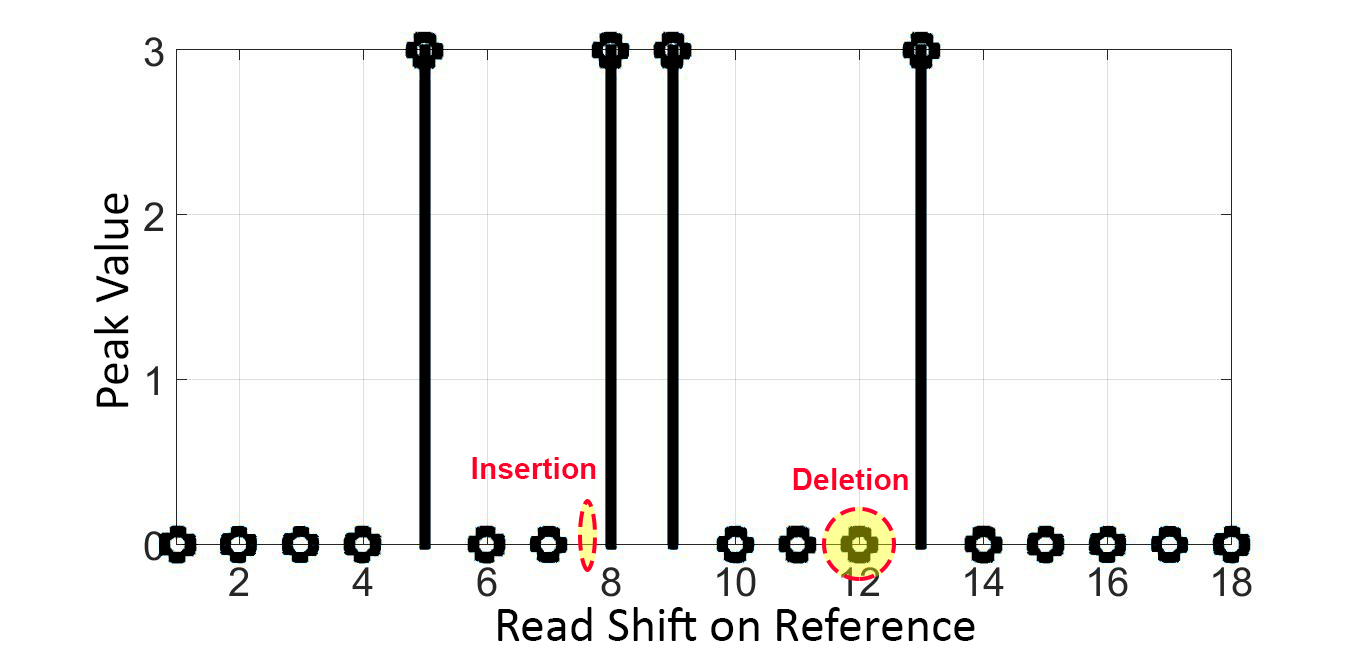}  
					\end{tabular}
				\end{center}
				\caption 
				{ \label{fig12}
					Full peaks in the read sequence for performing DNA sequence local alignment.} 
			\end{figure} 	
			Based on Fig. \ref{fig12}, it can be concluded that one character insertion has occurred in read sequence corresponding to the character between $7^\text{th}$ and $8^\text{th}$ character of the reference sequence. This is due in large to the fact that there is a full peak at $5^\text{th}$ character, so there is no edit at $5^\text{th}$, $6^\text{th}$, and $7^\text{th}$ character. On the other hand, there is a full peak at $8^\text{th}$ character, so the edit occurred prior to it. However, it should be mentioned that there is no edit at $7^\text{th}$ character. As a result, there is a character insertion between $7^\text{th}$ and $8^\text{th}$ characters. Moreover, similarly, we can demonstrate that there exists a character deletion in read sequence corresponding to the $12^\text{th}$ character of the reference sequence.
		\end{enumerate} 
		\subsection{Optical Implementation}
				\begin{figure}[!t]
					\begin{center}
						\begin{tabular}{c}
							\includegraphics[height=5.06cm,width=12.32cm]{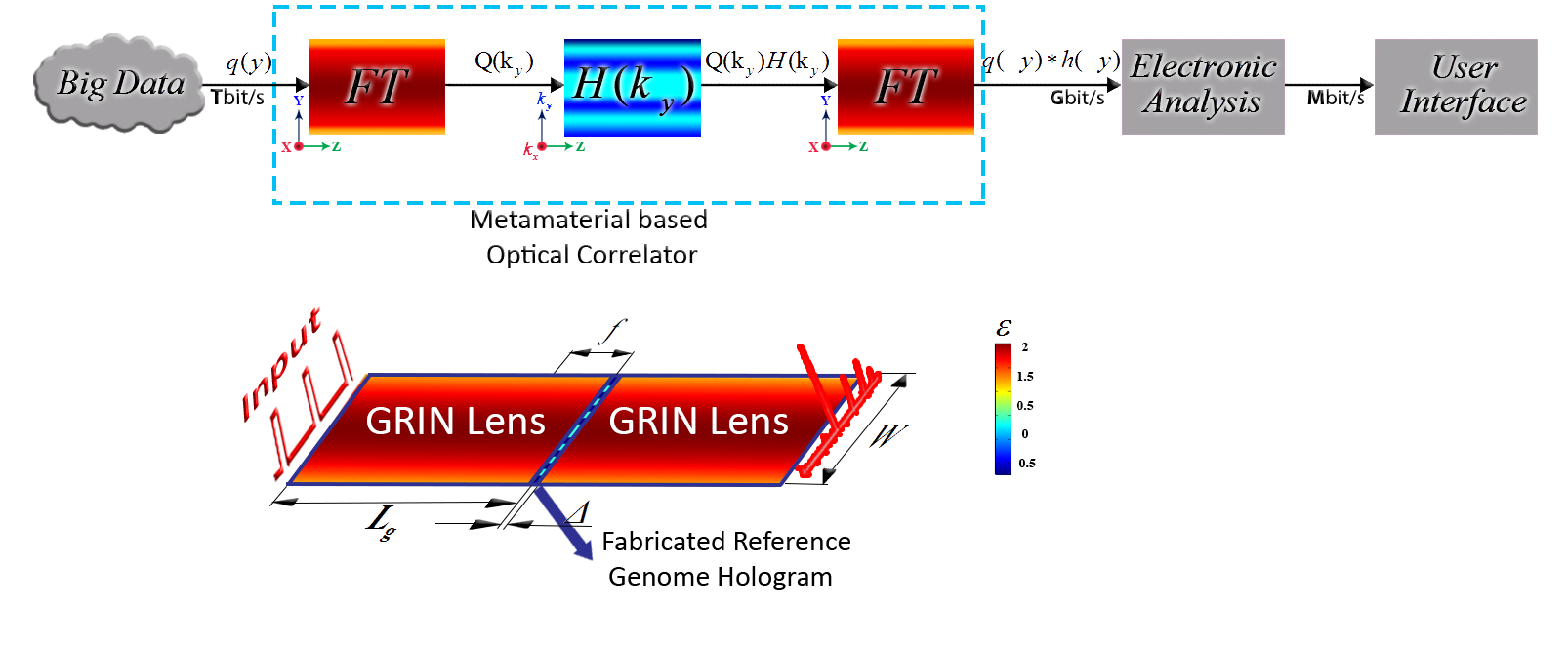} \\
							\includegraphics[height=4.62cm,width=13.2cm]{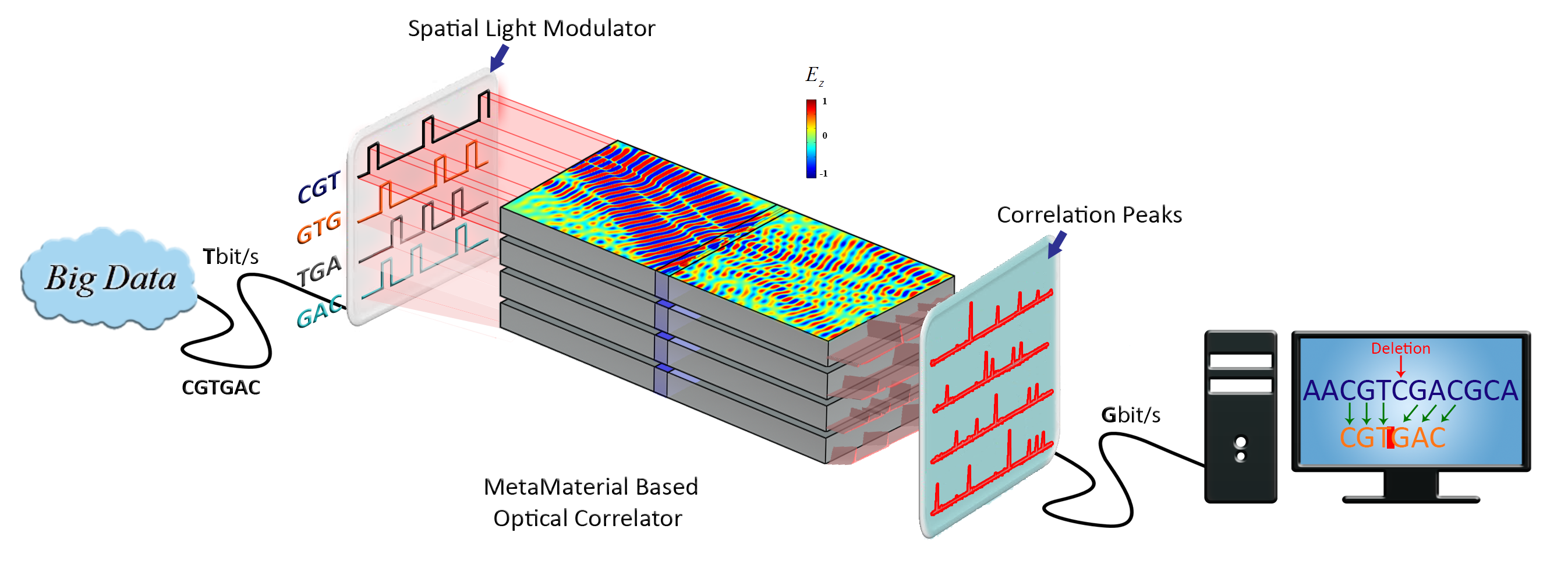}  
						\end{tabular}
					\end{center}
					\caption 
					{ \label{fig13}
						The proposed metamaterial optical correlator based on window size structure for DNA sequence alignment.} 
				\end{figure}
		For optical implementation, the reference genome should be stored on a hologram to be used as a fixed mask in the correlation process. Therefore, to implement an optical correlator for big data processing it is necessary to consider the size of the hologram, Fourier transform of the reference sequence, limitation. As the length of the reference genome supposed to be processed increases, its Fourier transform compresses on a smaller area on the hologram with more resolution due to space-bandwidth product principle. Therefore, high resolution hologram is needed for big data applications. Metamaterials with small a pixel size of $1$ nm have recently introduced paving the way for parallel processing \cite{32}. The structure of the proposed metamaterial-based optical correlator \cite{25} is shown in Fig. \ref{fig13}. As shown in this figure, reads namely queries are coded in the z-component of the electric field and then propagate through the graded index (GRIN) lens for producing their Fourier transform. On the other hand, Fourier transform of human genome as the reference genome is applied through the change in refractive index to a thin meta-material layer, named as meta-surface hologram. This Fourier transform once obtained for the reference genome is used in subsequent processing. The output pattern contains the multiplication of reference and queries Fourier transforms in each of its rows in the proposed structure which, after passing through another Fourier transform cylindrical GRIN lens, yields the cross-correlation of the reference genome and each query in each of its rows. The location of peaks in each row after some down-sampling determines the location of the corresponding query in the reference genome, and the exact location will be analyzed electronically as discussed in the previous section.
		To implement this optical correlator structure, some considerations should be taken to enable big data analysis. First, the system should be analyzed in terms of resolution and space-bandwidth product. The input query $q(y)$ is necessarily bounded within the transverse extent $W$. The first Fourier transform of $q(y)$ is likewise bounded within the following spatial frequency intervals \cite{33}:
		\begin{equation}
		\label{eq:4}
		\Delta {k_y} = \frac{W}{{{s^2}}}
		\end{equation}
		where, $s$ is the GRIN scale and is defined as follows:
		\begin{equation}
		\label{eq:5}
		s = \sqrt {\frac{{{\lambda _0}{L_g}}}{{{\pi ^2}\sqrt {\frac{\varepsilon }{{{\varepsilon _0}}}} }}}
		\end{equation}
		Where ${\lambda _0}$ is the incident wavelength, ${L_g}$ is the GRIN lens length, $\varepsilon$  and ${\varepsilon _0}$  are the vacuum and center GRIN lens permittivity, respectively. Boundedness in either domain implies that the meaningful information content of the signal in the other domain can be recovered from its samples at the Nyquist rate \cite{33}. Thus, the minimum resolvable dimension is approximately given by:
		\begin{equation}
		\label{eq:6}
		\delta {R_{res}} = \frac{W}{{\pi ({W^2}/{s^2})}} = \frac{{W{\lambda _0}{L_g}}}{{{\pi ^3}}}\sqrt {\frac{{{\varepsilon _0}}}{\varepsilon }} 
		\end{equation}
		which should be considered in the proposed optical correlator structure.
		Another important parameter that can limit the performance of the optical structure is the ability to fabricate a hologram with the highest possible resolution. Fabrication of a hologram with a $1$ nm resolution has been carried out recently \cite{32}. 
		
		Another parameter that determines the spectral resolution of the hologram, hologram pixel size, is computed as follows \cite{25}:
		\begin{equation}
		\label{eq:7}
		\delta {R_{hologram}} = \frac{W}{{{s^2}}} = \frac{{N{\lambda _0}{\pi ^2}\sqrt {\frac{\varepsilon }{{{\varepsilon _0}}}} }}{{{\lambda _0}{L_g}}} =  = \frac{{N{\pi ^2}\sqrt {\frac{\varepsilon }{{{\varepsilon _0}}}} }}{{{L_g}}}
		\end{equation}
		Finally, combining all the limiting spectral resolutions, the overall minimum limit of an optical correlator for the hologram pixel size can be calculated as:
		\begin{equation}
		\label{eq:8}
		\delta R = \max \{ \delta {R_{res}},\delta {R_{hologram}}\} 
		\end{equation}
		which should be considered while designing an optical correlator.
		In order to build a hologram corresponding to the reference human genome, Fourier transform of the genome with length $L_{genome}$ should be calculated as follows:
		\begin{equation}
		\label{eq:9}
		H({k_y}) = \int\limits_{ - \infty }^{ + \infty } {r(} y){e^{ - j2\pi {k_y}y}}dy
		\end{equation}
		Therefore, by equalizing ${e^{j\frac{{2\pi }}{{\lambda {}_0}}\Delta }}$  to the above equation and considering the GRIN lens scale, one can get the required permittivity as follows \cite{25}:
		\begin{equation}
		\label{eq:10}
		\frac{{{\varepsilon _{ms}}(y)}}{{{\varepsilon _0}}} = \frac{{{\mu _{ms}}(y)}}{{{\mu _0}}} = j\frac{{\lambda {}_0}}{{2\pi \Delta }}\ln [\frac{1}{{H(\frac{{{k_y}}}{{{s^2}}})}}]
		\end{equation}
		\section{Simulation Results}
		\label{sec33}
		In this section, a dataset of real human genome is used to evaluate speed and accuracy of the WOC method. Furthermore, in order to illustrate the superior performance of this method, compared to alternative sequence alignment techniques, the speed and accuracy of WOC is compared against the BLAST algorithm \cite{5}. The BLAST algorithm is the most well-known method, among different sequence alignment techniques proposed so far, due to its high speed and accuracy. 
		
		A portion of human DNA sequence obtained from PacBio Bioscience RS II platform \cite{13} is selected as the dataset. Table \ref{table2} summarizes characteristics of the used dataset.
		\begin{table}[!b]
			\caption{Dataset.} 
			\label{table2}
			\begin{center}       
				\begin{tabular}{|c|c|c|} 
					\hline
					\rule[-1ex]{0pt}{3.5ex}  Dataset & Platform & Read length (bps) \\
					\hline\hline
					\rule[-1ex]{0pt}{3.5ex}  Ref.\cite{13} & PacBio & 32 \\
					\hline
				\end{tabular}
			\end{center}
		\end{table} 
		In \cite{23}, statistical features of different occasions of edits in PacBio datasets are analyzed. As reported in this paper, (i) the sequencing error-rate is typically around $15\%$; (ii) errors have a uniform distribution around reads and do not occur at the end part of read sequences (this is also correct for other sequencing technologies); and finally (iii) $10\%$ of these errors are character insertions while $4\%$ and $1\%$ of them are in the form of character deletions and substitutions, respectively. 
		\subsection{Simulation Setup}
		In order to verify the correctness and accuracy of the proposed WOC method, we carried out several simulations. Simulation of the whole correlation process, depicted in Fig. \ref{fig6}, was performed in MATLAB, as well as, COMSOL Multiphysics as a physical simulator.

		The schematic of the physical structure implementing WOC method is sketched in Fig. \ref{fig13}. In our simulations, electric field wavelength (${\lambda _0}$) is $630$ nm. The GRIN lens has $128{\lambda _0}$  width and $54$ cm length. According to Equation (\ref{eq:7}), a resolution of $1.61$ nm is achieved for the designed hologram which is acceptable for a practical implementation \cite{32}. As the hologram length can be much smaller than the input light aperture, it would be rational to send query information via an spatial light modulator (SLM) and record reference sequence on a material as the hologram. Accordingly, the number of bits simultaneously coded on each raw of the input spatial light modulator is $128$ which equals to the GRIN length divided by ${\lambda _0}$. Therefore, $32$ characters can go through the structure in each row as the query, in our bit coding scheme.

		The holograms are selected in a way to hold $10000$ characters, thus, holograms of length $65$ $\mu m$ are placed in front of the Graded index lenses. Given that typical rate of error in read sequences is less than 15\%, the error rate of about 10\% is considered here. It is worth to mention that this error rate is the maximum error rate could be handled with BLAST algorithm. To this end, two insertions and a deletion is made in the query sequence. Thus, window size used for the correlation process of the proposed algorithm can be obtained from Equation (\ref{eq:2}). As the read length in our simulation is $32$ and the number of edits is $3$, the window size should be set to $8$ resulting in $25$ windows and correlation operations.
		
		\subsection{Results and Evaluation}
					\begin{figure}[!t]
						\begin{center}
							\begin{tabular}{c}
								\includegraphics[height=5.3cm,width=11cm]{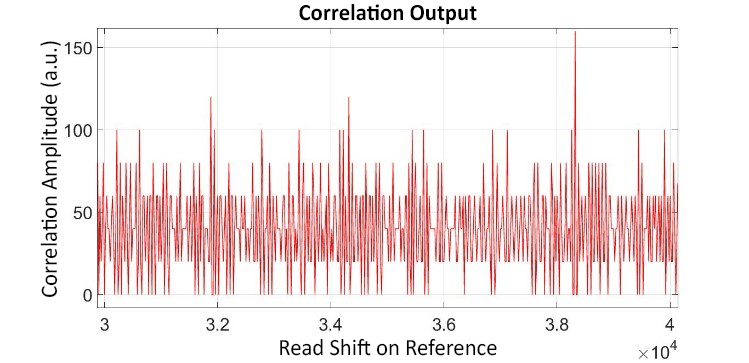}  
							\end{tabular}
						\end{center}
						\caption 
						{ \label{fig14}
							Optical Correlation output for the first window.} 
					\end{figure}
		\begin{figure}[!b]
			\begin{center}
				\begin{tabular}{c}
					\includegraphics[height=5.3cm,width=11cm]{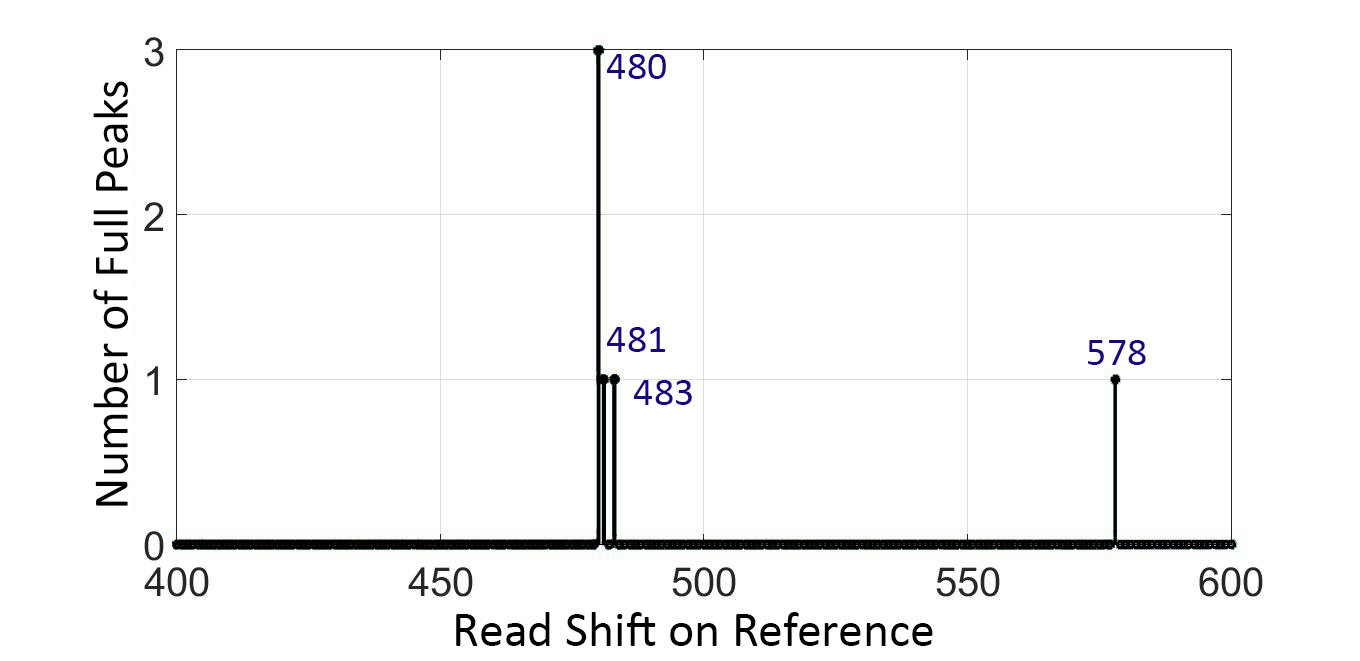}  
				\end{tabular}
			\end{center}
			\caption 
			{ \label{fig15}
				Number of full peaks to perform global alignment method, obtained through all windows, built upon the proposed metamaterial based optical correlator.} 
		\end{figure}
\begin{figure}[!t]
			\begin{center}
				\begin{tabular}{c}
					\includegraphics[height=5.3cm,width=11cm]{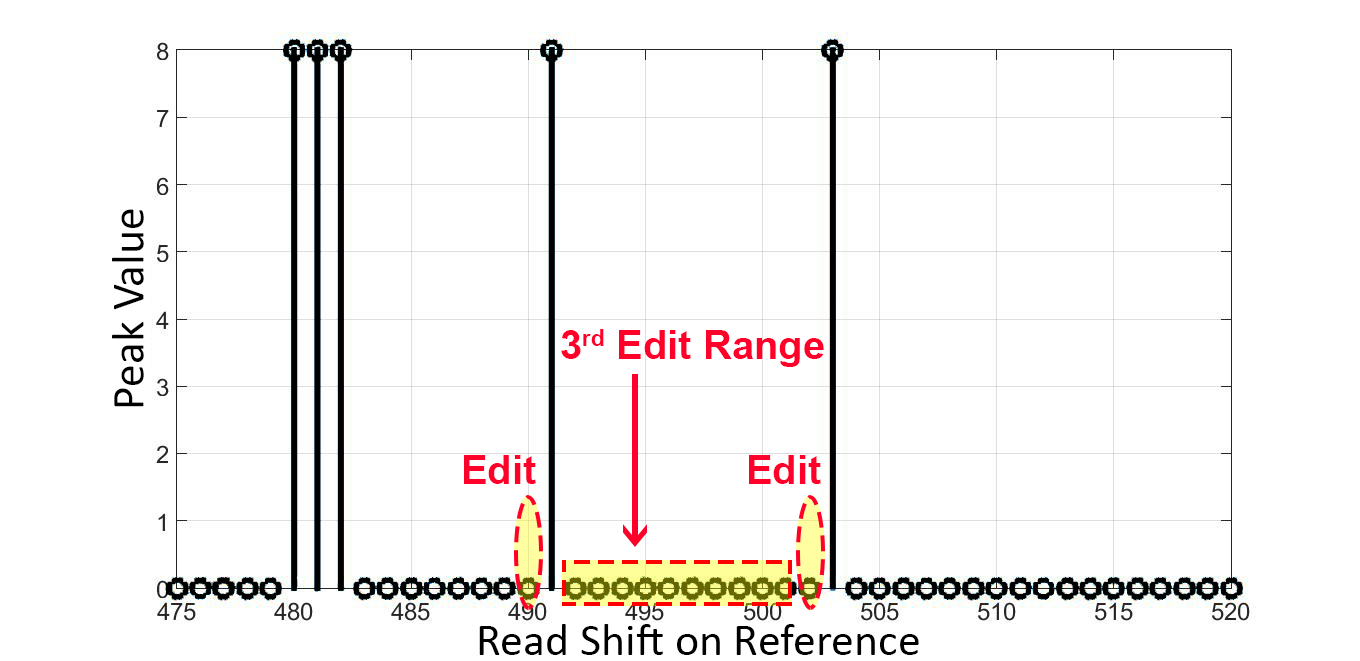}  
				\end{tabular}
			\end{center}
			\caption 
			{ \label{fig16}
				Correlation output of the proposed metamaterial based optical correlator for WOC analysis to exactly locate indels location through the proposed local alignment.} 
		\end{figure}

		Fig. \ref{fig14} illustrates the optical correlation result corresponding only to the first window to avoid redundant plots. The obtained correlation peaks from all windows are further used to perform global and local alignment. In order to analyze the correlation output, first, global alignment is conducted. As a result, approximate location of the read sequence is specified using the global alignment method described in Section \ref{sec24}. Then, all the correlation peaks in the range of read sequence location are considered. Fig. \ref{fig15} illustrates the number of full peaks obtained from the proposed structure. According to the peak locations and the correlation analysis of WOC method described in Section \ref{sec24}, local alignment can determine the exact locations of indels, as schematically shown in Fig. \ref{fig16}. As shown in this figure, although the exact locations of two edits are determined, the third edit is determined to be located in a sub-sequence represented by solid-red lines in Figure \ref{fig16}. This uncertainty to exactly locate the third indel is due to the large size of the window, against the small distance between two edits. In this case, two indels, located in a same window, cannot be detected. Therefore, smaller window size can solve the latter problem, as discussed in Section \ref{sec33}.
		It should be noted that since our coding scheme utilizes four bits for representing each character, it stands to reason that the correlation output should be down-sampled, in order to relax ambiguous matches resulted from codes overlap \cite{31}.

		\begin{figure}[!b]
			\begin{center}
				\begin{tabular}{c}
					\includegraphics[height=5.3cm,width=10.3cm]{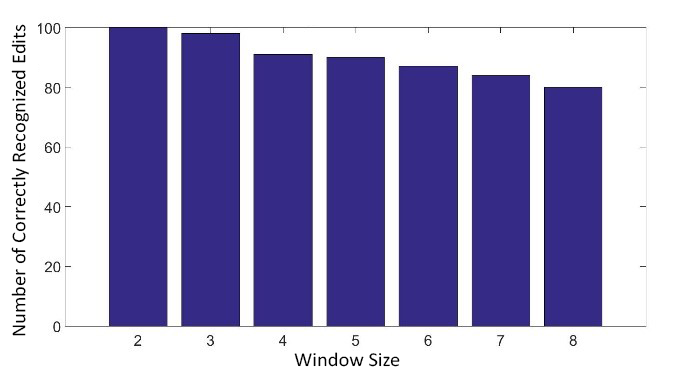}  
				\end{tabular}
			\end{center}
			\caption 
			{ \label{fig17after}
				Percentage of correctly located edits by WOC method for varying values of window size.} 
		\end{figure}   
		\subsection{Window Size Effect}
\begin{figure}[!t]
					\begin{center}
						\begin{tabular}{c}
							\includegraphics[height=5cm]{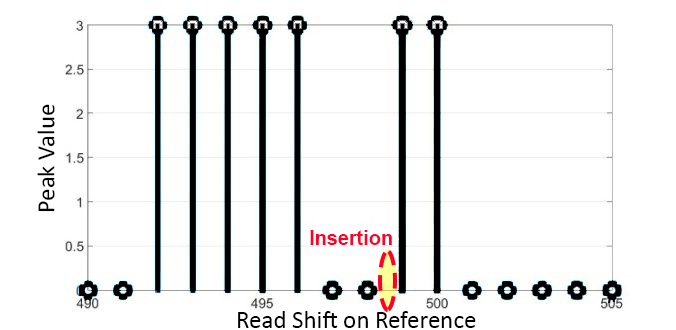}  
						\end{tabular}
					\end{center}
					\caption 
					{ \label{fig17}
						Successfully located edit by WOC method through local alignment with small window size.} 
				\end{figure}

		As is obvious, the selection of window size directly affects the accuracy of the proposed sequence alignment structure. Therefore, this study set out to assess the effect of window size on performance of the system. As can be seen in Fig. \ref{fig17after}, the proposed method can successfully locate 98\% of edits for $w_s= 3$. The accuracy of detection diminishes as the window size increases. As a result, we can adopt a larger window size (e.g. $w_s= 8$) for finding the place of read, and then, in the case that the exact locations of all edits are not detected, we can adopt smaller window size (e.g. $w_s= 2$ or $3$) to exactly locating the edits. In this manner, the region, through which WOC method locates the edits, becomes smaller.
		\begin{algorithm}
			
			\SetKwData{Left}{left}
			\SetKwData{This}{this}
			\SetKwData{Up}{up}
			\SetKwFunction{Union}{Union}
			\SetKwFunction{FindCompress}{FindCompress}
			\SetKwInOut{Input}{input}
			\SetKwInOut{Output}{output}
			\SetKwInOut{Functions}{Functions}
			\SetKwInOut{Pseudocode}{Pseudocode}
			\Input{$L_R$ (read length), $e$ (\# of edits), read (read sequence), ref (reference sequence)}
			\Output{exact location of $e$ edits}
			\Functions{read\_partitioning: partitioning read sequence into n overlapped-windows by the size of $w_s$ \\
				global\_alignment: determining boundary of high similarity by \# of full peaks\\ from windows correlation outputs \\
				local\_alignment: detecting exact location of edits in high similarity boundary \\
				detecting\_edit\_in\_small\_area: determining exact location of remained edits in a\\ small area of read and reference sequence, by minimizing window size to 3 \\}
			\Pseudocode{}
			\label{Al1}			
			\BlankLine
			\tcp{Calculating window size and \# of windows}
			${w_s} = \left\lfloor {\frac{{{L_R}}}{{e + 1}}} \right\rfloor $ \\
			$n = {L_R} - {w_s} + 1$\\
			\tcp{Partitioning read sequence}
			readWindows[n] = read\_partitioning (read, $w_s$);\\
			\tcp{Correlating reference sequence by n windows of read sequence in parallel}
			correlationOutputs[n] = ref() $\otimes$ readWindows[n];\\
			\tcp{Global alignment: Calculating \# of full peaks from all windows correlation outputs}
			\For{$k\leftarrow 1$ \KwTo $n$}{
				correlationOutputs[k] $\leftarrow$ correlationOutputs[k]$<<$(k-1);\\
				\tcp{whole peak of read sequence is composed of 1st peak of window1, 2nd peak of window2,...., nth peak of window n}}
			\For{$i\leftarrow 1$ \KwTo $size(ref())$}{	\For{$j\leftarrow 1$ \KwTo $n$}{\label{forins}
					\If(){	correlationOutputs[j] == $w_s$}{\label{lt}
						numberOfFullPeaks[i] = numberOfFullPeaks[i] +  1;\\
					}}}
					highSimilarityBoundary =global\_alignment[numberOfFullPeaks];\\
					\tcp{Local alignment: detecting exact location of edits}
					editLocations = local\_alignment(highSimilarityBoundary,$w_s$, correlationOutputs);\\
					\If(){size(editLocations)$<$e}{\label{ut}
						$w_s$=3;\\			
						editLocations.pushback(detecting\_edit\_in\_small\_area());\\}
					\textbf{return} editLocations;
					
					\caption{Edit Detection Pseudo-code by Correlation Results}\label{al1}
				\end{algorithm}
				In previous example, third edit was detected in the range of [492,501]. Therefore, by decreasing window size into three, the exact location of third edit can be detected. Correlation output of performing local sequence alignment for $w_s= 3$ is shown in Fig. \ref{fig17} for an appropriate region.
				Summing up, the proposed window based correlation algorithm for detecting exact locations of edits is summarized as pseudo code in Algorithm \ref{Al1}.
				
				\subsubsection{Algorithm Comparison with BLAST}
 
				In this section, results of several simulations comparing performance of the WOC method with BLAST algorithm are presented.
				
				The same set of queries and databases, discussed in previous section, is used in simulating both WOC and BLAST algorithms to compare their accuracy and run time.  Note that as the read sequence is a short string with 32 bps long, the simulation is performed at the presence of 1, 2, 3, 4, 5 and 6 edits. To analyze the accuracy of the system, following equation is utilized  \cite{22}:
								\begin{figure}[!t]
									\begin{center}
										\begin{tabular}{c}
											\includegraphics[height=5cm]{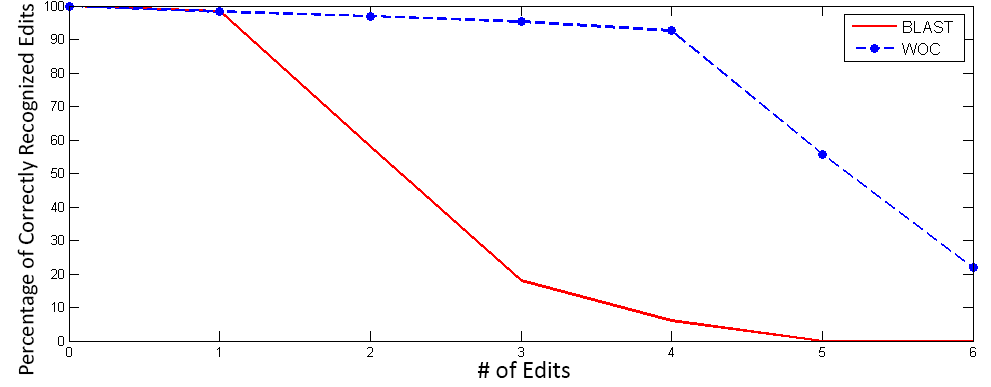}  
										\end{tabular}
									\end{center}
									\caption 
									{ \label{fig19}
										Accuracy comparison of the proposed WOC with BLAST.} 
								\end{figure}
				\begin{equation}
				\label{eq:100}
				Accuracy = \frac{True~Positive}{True~Positive + False~Positive}
				\end{equation}
				As it can be seen from Fig. \ref{fig19}, WOC method and BLAST algorithms show similar accuracy in the case of low number of edits, but as the number of edits increases, BLAST sensitivity rapidly declines. However, WOC method preserves accuracy even at the presence of high number of edits.  This high accuracy is obtained through decreasing the required window size for detecting all edits. Therefore, our structure is sensitive to very high number of edits, while BLAST is not able to perform DNA sequence alignment in the case of high number of edits, foe example for number of edits more than $3$ bps.

\begin{figure}[!t]
									\begin{center}
										\begin{tabular}{c}
											\includegraphics[height=5cm]{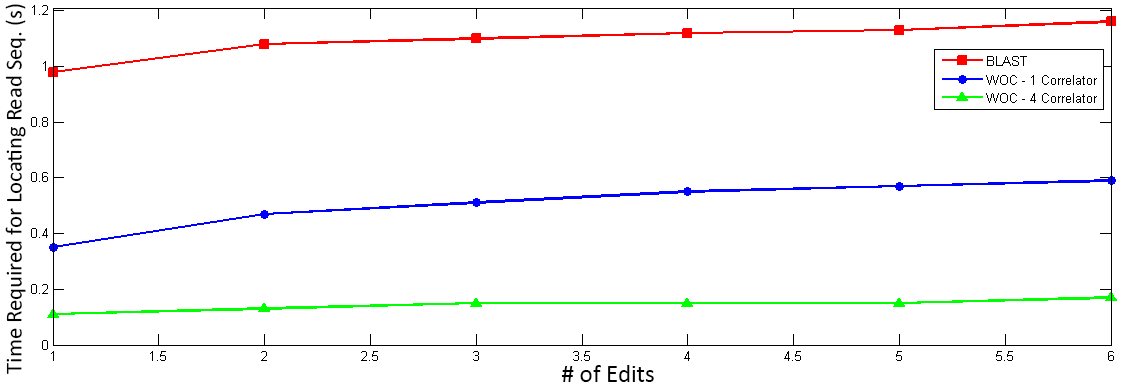}  
										\end{tabular}
									\end{center}
									\caption 
									{ \label{fig20}
										Time comparison of the proposed WOC with BLAST.} 
								\end{figure}

				Moreover, a comparison for run time, in seconds, among WOC and BLAST methods, is made at the presence of different number of edits. In this manner, read sequence is searched in a reference sequence of $10000$ character length. Run time taking by WOC method is the sum of time intervals required for performing optical correlations, as well as the consumed time to perform peak analysis electronically. The correlation time depends on the number of correlators performed in parallel for different windows. For the number of correlators chosen closer to the number of windows, parallelization becomes more efficient, and the corresponding run time will be lower. To evaluate the required time by optical correlation part, first, we assume the number of correlators to be four, and then, only one correlator is utilized to account for the worst case. Note that each correlation approximately takes $20$ ms, including the spatial light modulator switching time and data acquisition \cite{37,32}. In addition, peaks analysis is performed by a typical electronic based computer with $3$ GHz CPU clock. This time is computed by pseudo code of Algorithm \ref{Al1}. Run time of this algorithm is on $O(n\times size(reference))$. Evaluation results of WOC-1 correlator and WOC-4 correlator are shown in Fig. \ref{fig20} for different numbers of edits. On the other hand, required time for the well-known BLAST algorithm is extracted for the corresponding simulation. As shown in Fig. \ref{fig20}, BLAST takes more than one second to perform sequence alignment. However, it is crystal clear from Fig. \ref{fig20} that time required for WOC method is much lower than BLAST algorithm.			 
				\section{Conclusion}
				In this paper, an optical correlation based algorithm able to efficiently compare two spatially coded DNA sequences was proposed. The proposed structure showed high reliability even in the case that the query sequence has a high number of edits for global alignment.  Moreover, we proposed an accurate correlation based algorithm to perform local alignment that benefited from optical parallel processing. Global and Local alignments were achieved by partitioning the query to a number of windows and performing simultaneous correlations for each partition. The correlation results were further analyzed by a simple algorithm; accordingly, to exactly locate the indels. To evaluate the performance of the window based optical correlation method (named as WOC) proposed for DNA sequence alignment, a comparison was also made with the well-known electronic based alignment algorithm, i.e. BLAST algorithm. Comparing the corresponding results, the proposed optical correlator leads to better performance and higher accuracy in finding queries and locating edits, especially when the number of edits in the genome increases. Although, WOC algorithm was faster in performing DNA sequence alignment, with regard to BLAST, still the unlimited computation parallelism  inherent by optical computing can be used to provide a much superior performance of optical computing in this problem. Therefore, taking advantages of parallel optical computing, the proposed optical correlation-based sequence alignment structure reveals huge reduction in the processing time, compared to electronic based architectures such as BLAST.
				\subsection*{Disclosures}
The authors have no relevant financial interests in this article and no potential conflicts of interest to disclose.
				\bibliographystyle{spiejour}
				\bibliography{Ref}
								\listoffigures
								\listoftables
			\end{spacing}
		\end{document}